\pdfoutput=1
\documentclass[authoryear,12pt]{elsarticle}

\usepackage[T1]{fontenc}
\usepackage[latin9]{inputenc}
\usepackage[active]{srcltx}
\usepackage{array}
\usepackage{mathtools}
\usepackage{amsbsy}
\usepackage{amssymb}
\usepackage{mathdots}
\usepackage{graphicx}
\usepackage{esint}

\newcommand{\apj}{ApJ}

\newcommand{\pre}{Phys. Rev. E}
\newcommand{\apjl}{ApJL}
\newcommand{\apjs}{ApJ Supplement}

\newcommand{\aap}{A~\&~A}

\providecommand{\tabularnewline}{\\}

\usepackage[a4paper=true,%
breaklinks=true,%
colorlinks=true,%
pdfauthor={First Author et al.},%
pdftitle={for special issue  of Advances in Space Research, "Dynamo Frontiers" }%
]{hyperref}

\journal{Advances in Space Research}

\begin{document}

\begin{frontmatter}

\title{Angular momentum fluxes caused by $\Lambda$-effect and meridional circulation
structure of the Sun}

\author{Valery Pipin\corref{cor} }
\address{Institute of Solar-Terrestrial Physics, Russian Academy of
Sciences, Irkutsk, 664033, Russia}
\cortext[cor]{Corresponding author}
%\fntext[footnote2]{Additional information regarding the corresponding author}
\ead{pip@iszf.irk.ru}

% Url can be given like this:
% \ead[url]{http://www.elsevier.com/wps/find/authorsview.authors/latex}
\author{Alexander Kosovichev}
\address{New Jersey Institute of Technology, Newark, NJ 07102, USA}
%%%\fntext[footnote3]{Additional information about the second and third authors}
%%%\ead{sasha@bbso.njit.edu}

\begin{abstract}
Using mean-field hydrodynamic models of the solar angular momentum balance we
show that the non-monotonic latitudinal dependence of the radial angular momentum
fluxes caused by $\Lambda$-effect  can affect the number of the
meridional circulation cells stacking in radial direction in the solar
convection zone. In particular, our results show the possibility of a complicated
triple-cell meridional circulation structure. This pattern consists of
two large  counterclockwise circulation
cells (the N-hemisphere) and a smaller clockwise cell located
at low latitudes at the bottom of the convection zone.
\end{abstract}

\begin{keyword}
%first keyword \sep second keyword \sep more keywords
Sun; differential rotation; meridional circulation
% PACS codes here, in the form:
\PACS 96.60.Q\sep 96.60.Jw
\end{keyword}

\end{frontmatter}

\section{Introduction}

The mean-field models have been  successful in explaining basic properties
of the differential rotation of the Sun and solar-like stars, e.g.,
\cite{kuk05} and \cite{kit11}.
These models predicted dependence of the the surface
latitudinal shear on the rotation rate and the spectral class for a
set of low-main-sequence solar-like stars with external convection
zone, \citep{1999AA...344..911K}. However, recent findings from
analysis of the supergranulation
dynamics on the Sun \citep{hath12} and helioseismology inversions \citep{Zhao13m,khol2014}
put in question the results of the mean-field models predicting a
single-cell meridional
circulation structure of the solar convection zone. Global 3D simulations
of the flows in the solar convection zone also show the multiple-cell
structure of the meridional circulation \citep{kap2011,miesch11,guer2013}. 

In this note we show that the double-cell meridional circulation structure
can be compatible with the solar-like rotation law in the framework of
the mean-field theory. The $\Lambda$-effect is one of the key
ingredient of the mean-field models of solar differential rotation
It is caused by effect of the Coriolis forces on convective  flows
 the stratified rotating  convection zones \citep{KR93L}.  The
 $\Lambda$-effect produces non-disspative angular momentum
 fluxes. Therefore the spatial structure of the  $\Lambda$-effect
 determine the spatial structure of the global flows in the stellar
 convection zone. For example, the latitudinal dependence of the  $\Lambda$-effect
determines the radial profile of the angular velocity
distribution  \citep{KR93L}. Here, we study how it affects the structure of the
meridional circulation.  
The study is carried
out using the standard mean-field models of the angular momentum balance
and heat transport in the solar convection zone.

\section{Basic equations}

The reference internal thermodynamic structure of the Sun is calculated
using the MESA stellar evolution code (version r7623)
\citep{mesa11,mesa13}. The model is calculated using the mixing length
parameter  $\alpha_{MLT}={\displaystyle \frac{\ell}{H_{p}}}=2$,
where $H_{p}$ is the pressure scale.

\subsection{Heat transport}

Following to \citet{1999AA...344..911K}, effects of rotation
on the thermal balance are calculated from the mean-field heat
transport equation,
\begin{equation}
\overline{\rho}\overline{T}\frac{\partial s}{\partial t}+\overline{\rho}\overline{T}\left(\overline{\mathbf{U}}\cdot\boldsymbol{\nabla}\right)s=-\boldsymbol{\nabla}\cdot\left(\mathbf{F}^{conv}+\mathbf{F}^{rad}\right),\label{eq:heat}
\end{equation}
 where  $\mathbf{\overline{U}}$
is axisymmetric mean flow. We employ the expression for the anisotropic
convective flux suggested by \citet{kit-pip-rud},
\begin{equation}
F_{i}^{conv}=\overline{\rho}\overline{T}\chi_{ij}\nabla_{j}s,\label{conv}
\end{equation}
where the heat conductivity tensor $\chi_{ij}$ reads
\[
\chi_{ij}=\chi_{T}\left(\phi\left(\Omega^{*}\right)+c_{\chi}\phi_{\parallel}\left(\Omega^{*}\right)\frac{\Omega_{i}\Omega_{j}}{\Omega^{2}}\right).
\]
Functions $\phi$, $\phi_{\parallel}$ are defined in the above
cited paper. The effect of the global rotation on the heat transport
depends on the Coriolis number $\Omega^{*}=2\tau_{c}\Omega_{0}$,
where $\tau_{c}$ is the turn-over time of convective flow. Following
to \citet{kit11} we assume $c_{\chi}=1.5$. If we neglect rotation ($\Omega^{*}\rightarrow0$), the heat conductivity tensor
reduces to the standard form $\chi_{ij}={\displaystyle
  \frac{1}{3}\delta_{ij}\tau_{c}}u'^{2}$, where $u'$ is the
RMS convective velocity, which is determined from the mixing-length
relation 
\[
u'^{2}=-\frac{\ell^{2}g}{4c_{p}}\frac{\partial s}{\partial r}.
\]
Thus, the turbulent heat conductivity is
\begin{equation}
\chi_{T}=-\frac{\tau_{c}\ell^{2}g}{12c_{p}}\frac{\partial s}{\partial r},\label{eq:chit}
\end{equation}
The radiative heat transport is:,
\[
\mathbf{F}^{rad}=-c_{p}\overline{\rho}\chi_{D}\boldsymbol{\nabla}T,
\]
 where
\[
\chi_{D}=\frac{16\sigma\overline{T}^{3}}{3\kappa\overline{\rho}^{2}c_{p}},
\]
where $\kappa$ is the opacity coefficient. The radial profiles of the
gravity acceleration, $g$, the density, $\overline{\rho}$, the temperature,
$\overline{T}$,  as well as others parameters: $c_{p}$,  $\kappa$ ,
$\tau_{c}$ and  $\ell$ are estimated from  MESA code. The integration domain of the mean-field
model is from $r_{i}=0.715R_{\odot}$ to $r_{e}=0.99R_{\odot}$. At
the inner boundary the energy flux in radial direction is
$F_{r}^{conv}+F_{r}^{rad}={\displaystyle
  \frac{L_{\odot}\left(r_{i}\right)}{4\pi r_{i}^{2}}}$, 
and at the outer boundary, following to \citet{kit11}, we
apply: 
\[
F_{r}=\frac{L_{\odot}}{4\pi r_{e}^{2}}\left(1+\left(\frac{s}{c_{p}}\right)^{4}\right).
\]

\subsection{Angular momentum balance}

The heat transport equation is coupled to equations for the angular
momentum balance. This balance is governed by the conservation of
the angular momentum \citep{1989drsc.book.....R}. In the spherical
coordinate system it is expressed as follows: 
\begin{equation}
\frac{\partial}{\partial t}\overline{\rho}r^{2}\sin^{2}\theta\Omega=-\boldsymbol{\nabla\cdot}\left(\overline{\rho}r\sin\theta\left(\hat{\mathbf{T}}_{\phi}+r\sin\theta\Omega\mathbf{\overline{U}^{m}}\right)\right),\label{eq:az}
\end{equation}
where the mean flow satisfies the continuity equation, 
\begin{equation}
\boldsymbol{\nabla}\cdot\overline{\rho}\mathbf{\overline{U}}=0,\label{eq:cont}
\end{equation}
where $\mathbf{\overline{U}}=\mathbf{\overline{U}}^{m}+r\sin\theta\Omega\hat{\mathbf{\boldsymbol{\phi}}}$
and $\boldsymbol{\hat{\phi}}$ is the unit vector in the azimuthal direction.
The equation for the azimuthal component of the vorticity of the large-scale
flow,
$\omega=\left(\boldsymbol{\nabla}\times\overline{\mathbf{U}}^{m}\right)_{\phi}$ ,
is 
\begin{align}
\frac{\partial\omega}{\partial t} & =-\left[\boldsymbol{\nabla}\times\frac{1}{\overline{\rho}}\boldsymbol{\nabla\cdot}\overline{\rho}\hat{\mathbf{T}}\right]_{\phi}+r\sin\theta\frac{\partial\Omega^{2}}{\partial z}+\frac{1}{\overline{\rho}^{2}}\left[\boldsymbol{\nabla}\overline{\rho}\times\boldsymbol{\nabla}\overline{p}\right]_{\phi},\label{eq:vort}
\end{align}
where $\partial/\partial z=\cos\theta\partial/\partial r-\sin\theta/r\cdot\partial/\partial\theta$
is the gradient along the axis of rotation,  $\hat{\mathbf{T}}$ is
the turbulent part of the stresses, which is determined from the mean-field
hydrodynamics theory (see, \citealp{KR93L,kit-pip-rud}) as follows 
\begin{equation}
\hat{T}_{ij}=\left\langle u_{i}u_{j}\right\rangle,\label{eq:stres}
\end{equation}
where $\mathbf{u}$  is  fluctuating velocity. 
The first term in the RHS of the Eq.(\ref{eq:vort})
describes dissipation of the mean vorticity, $\omega$. Similarily
to \citet{rem2005ApJ} we approximate it as follows, 
\begin{equation}
-\left[\boldsymbol{\nabla}\times\frac{1}{\overline{\rho}}\boldsymbol{\nabla\cdot}\overline{\rho}\hat{\mathbf{T}}\right]_{\phi}\approx\nu_{T}\phi_{1}\nabla^{2}\omega,\label{eq:Re}
\end{equation}
where $\nu_{T}={\displaystyle \frac{4}{5}\chi_{T}},$ and the function
$\phi_{1}\left(\Omega^{*}\right)$ takes into account effects
of rotation on the turbulent viscosity \citep{kit-pip-rud}.
For the ideal gas the last term in Eq.(\ref{eq:vort}) can be rewritten
in terms of the specific entropy \citep{1999AA...344..911K}, 
\begin{equation}
\frac{1}{\overline{\rho}^{2}}\left[\boldsymbol{\nabla}\overline{\rho}\times\boldsymbol{\nabla}\overline{p}\right]_{\phi}\approx-\frac{g}{r
  c_{p}}\frac{\partial s}{\partial\theta}.\label{eq:baroc}
\end{equation}
The meridional circulation velocity $\overline{\mathbf{U}}^m$ is
expressed via stream function $\Psi$ :
$\overline{\mathbf{U}}^{m}={\displaystyle
  \frac{1}{\overline{\rho}}}\boldsymbol{\nabla}\times\Psi$, and,
\begin{equation}
-\overline{\rho}\omega=\left(\Delta-\frac{1}{r^{2}\sin^{2}\theta}\right)\Psi-\frac{1}{r\overline{\rho}}\frac{\partial\overline{\rho}}{\partial r}\frac{\partial r\Psi}{\partial r}.\label{eq:poiss}
\end{equation}

We employ the stress-free boundary conditions for Eq.(\ref{eq:az}),
the azimuthal component of the mean vorticity, $\omega$, is put to
zero at the boundaries.

The turbulent  angular momentum flux is \citep{1999AA...344..911K}: 
\begin{eqnarray}
T_{r\phi} & = & \nu_{T}\left\{ \psi_{\perp}+\left(\psi_{\parallel}-\psi_{\perp}\right)\mu^{2}\right\} r\frac{\partial\sin\theta\Omega}{\partial r}\nonumber \\
 & + & \nu_{T}\sin\theta\left(\psi_{\parallel}-\psi_{\perp}\right)\left(1-\mu^{2}\right)\frac{\partial\Omega}{\partial\mu}\label{eq:trf}\\
 & - & \nu_{T}\sin\theta\Omega\left(\frac{\ell}{H_{\rho}}\right)^{2}\left(V^{(0)}+\sin^{2}\theta V^{(1)}\right),\nonumber \\
T_{\theta\phi} & = & \nu_{T}\sin^{2}\theta\left\{ \psi_{\perp}+\left(\psi_{\parallel}-\psi_{\perp}\right)\sin^{2}\theta\right\} \frac{\partial\Omega}{\partial\mu}\nonumber \\
 & + & \nu_{T}\left(\psi_{\parallel}-\psi_{\perp}\right)\mu\sin^{2}\theta r\frac{\partial\Omega}{\partial r}\label{eq:ttf}\\
 & - & \nu_{T}\mu\Omega\sin^{2}\theta\left(\frac{\ell}{H_{\rho}}\right){}^{2}H^{(1)},\nonumber 
\end{eqnarray}
where $\nu_{T}={\displaystyle Pr_{T}\chi_{T}}$, $\mu=\cos\theta$.
The mean-field theory gives the turbulent Prandtl number
$Pr_{T}=\frac{4}{5}$ \citep{kit-pip-rud}, but  we consider $Pr_{T}$ as a free parameter. The viscosity
functions: $\psi_{\|},\psi_{\perp}$, and the $\Lambda$- effect
parameters $V^{\left(0,1\right)}$ and $H^{\left(0,1\right)}$  depend
on the Coriolis number and anisotropy parameters. 
In this paper we assume that 
 $V^{(1)}=V_{0}^{(1)} (1-a f(r))$, where $V_{0}^{(1)}=0.1$ 
is suggested by \citet{KR93L} for the case of the fast rotation,
$\Omega^{*}\gg1$. The adhoc function $f(r)$ models the subsurface rotational
shear layer and $a$ is the anisotropy parameter. Calculations of
Kitchatinov \& Ruediger (2005) suggest that the effect of the
prescribed anisotropy of convective flows on the non-disspative
angular momentum flux is strongly reduced in the deep layers of the
Sun. We model this as follows 
\begin{equation}
f(r)=\frac{1}{2}\left(1-erf\left(50\left(x_{s}-\frac{r}{R_{\odot}}\right)\right)\right)\label{eq:fr}
\end{equation}
This equation shows that the anisotropy effect,
which is controlled by parameter $a$, is restricted to external layer
of the convection zone, i.e., the layer that is above of the $x_{s}=0.95R_{\odot}$.
The other components of the $\Lambda$ effect are
parametrized as follows 
\begin{equation}
V^{(0)}=-V_{0}^{(1)} c_{\Lambda},\label{eq:V}
\end{equation}
and $H^{(1)}=V^{(1)}$. 

Let's summarize the basic assumptions of our model. The
reference thermodynamic structure of the solar convection zone is
computed from the stellar evolution code (MESA, r7623) for the non-rotating
Sun of age 4.6Gyr.  Eq(\ref{eq:heat}) governs deviations of
the entropy distribution from the reference state due to global
flows. The global flows are determined from the angular momentum
balance,  taking into account of sources of the meridional circulation
due to imbalance of the centrifugal and baroclinic forces. The Taylor
number in the model determines the strength of the centrifugal forces,
\begin{equation}
Ta=\frac{4\Omega_{0}^{2}R_{\odot}^{4}}{\nu_{T}^{2}},\label{eq:Ta}
\end{equation}
For the theoretical turbulent Prandtl number, $Pr_{T}=\frac{4}{5}$,
the magnitude of the eddy diffusivity is $\nu_{T}\approx10^{13}$cm$^{2}$/s,
and $Ta\sim8\cdot10^{6}$ . 
Using parameters $V_{0}^{(1)}=0.1$,
$c_{\Lambda}=0.9$ and 
$a=1.35$ the model reproduces the
angular velocity distribution in agreement  with solar
observation, and predicts  the one-cell meridional circulation
structure. Parameters of the model which we use in the numerical
experiments are listed in the Table 1 

\begin{table}
\protect\caption{Parameters of the models}

\centering{}%
\begin{tabular}{|>{\raggedright}p{2.5cm}|l|>{\raggedright}p{2.5cm}|}
\hline 
Model & $c_{\Lambda}$ & $Pr_{T}$\tabularnewline
\hline 
M1 & 0.9 & $1/3$\tabularnewline
\hline 
M2 & 5/6 & $1/2$\tabularnewline
\hline 
M3 & 3/4 & $1/2$\tabularnewline
\hline 
\end{tabular}
\end{table}

\section{Results}

Figure 1 and 2 illustrate the distribution of the turbulent parameters
and the angular velocity for  model M1. The angular velocity profile is in agreement
with the helioseismology results of \citet{Howe2011JPh}. The model shows one counterclockwise
circulation cell in the Northern hemisphere with the  amplitude of the
flow velocity 
about 10 m/s at the surface and at the bottom of the convection zone.
This model is in agreement with results of \citet{kit11}. 

\begin{figure}
\includegraphics[width=0.95\linewidth]{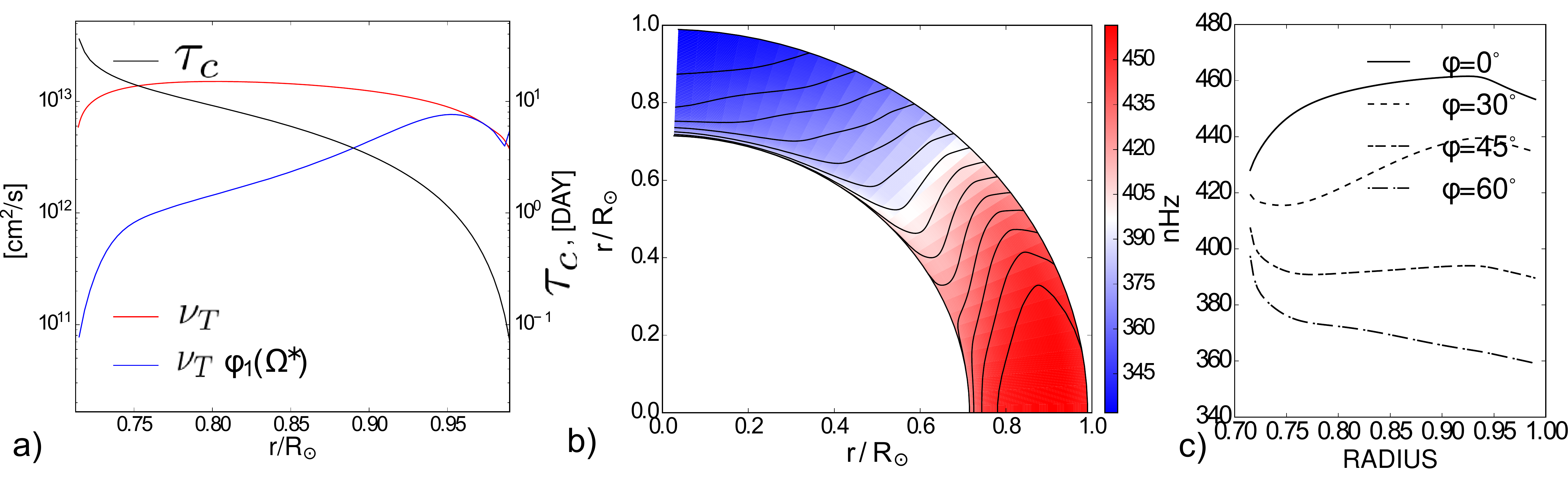}

\protect\caption{\label{fig:a)-The-output} Convection zone
  properties from the MESA solar model: a) the convective
turnover time, $\tau_{c}$ and the turbulent diffusivity parameter 
(red line), isotropic eddy viscosity  is shown by blue line;
 b) the angular velocity profile of model M1 in the
Northern hemisphere; c) the radial profile of the angular velocity
for the different latitudes.}
\end{figure}

\begin{figure}
\includegraphics[width=0.95\linewidth]{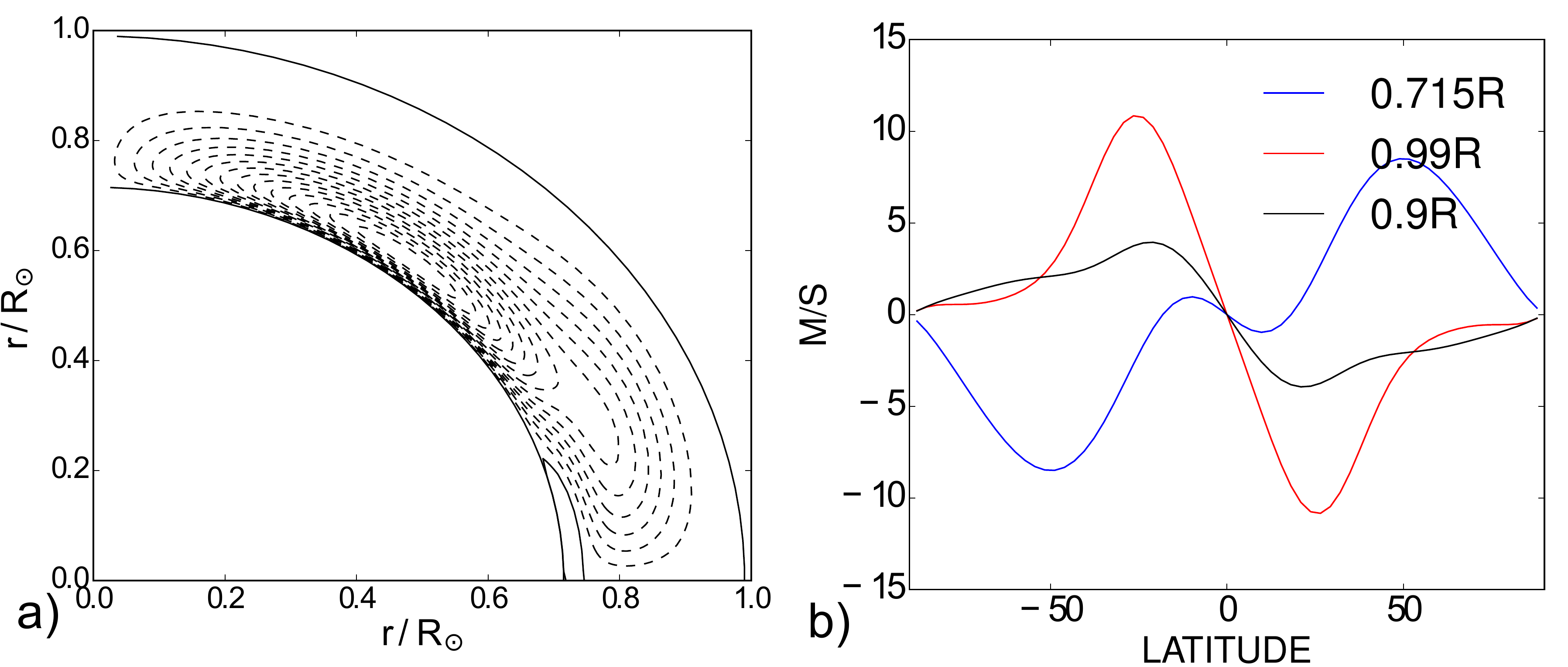}

\protect\caption{\label{fig:M1circ}Model M1. a) Streamlines of the meridional flow (
stream function $\Psi$) shows a single-cell counterclockwise circulation pattern
for the Northern hemisphere; b) velocity of the latitudinal component
of the meridional flow for different radii.}
\end{figure}

In model M2  parameter $c_{\Lambda}$ is smaller than  in model
M1. The increase of the inward angular momentum flux due to the decrease of
 $c_{\Lambda}$ redistributes of the non-dissipative angular momentum
fluxes. This results in an increase of  the latitudinal shear. To compensate
this effect we increase the Prandtl number from $1/3$ to $1/2$ (see, Table 1). The
resulted pattern of the meridional flow has a triple-cell structure
shown in Figure 3. This pattern consists of the two large counterclockwise
circulation cells and  a  small clockwise cell
located at low latitude at the bottom of the convection zone. The upper
equatorial circulation cell has a stagnation point at the $r=0.86R$. 

\begin{figure}
\includegraphics[width=0.95\linewidth]{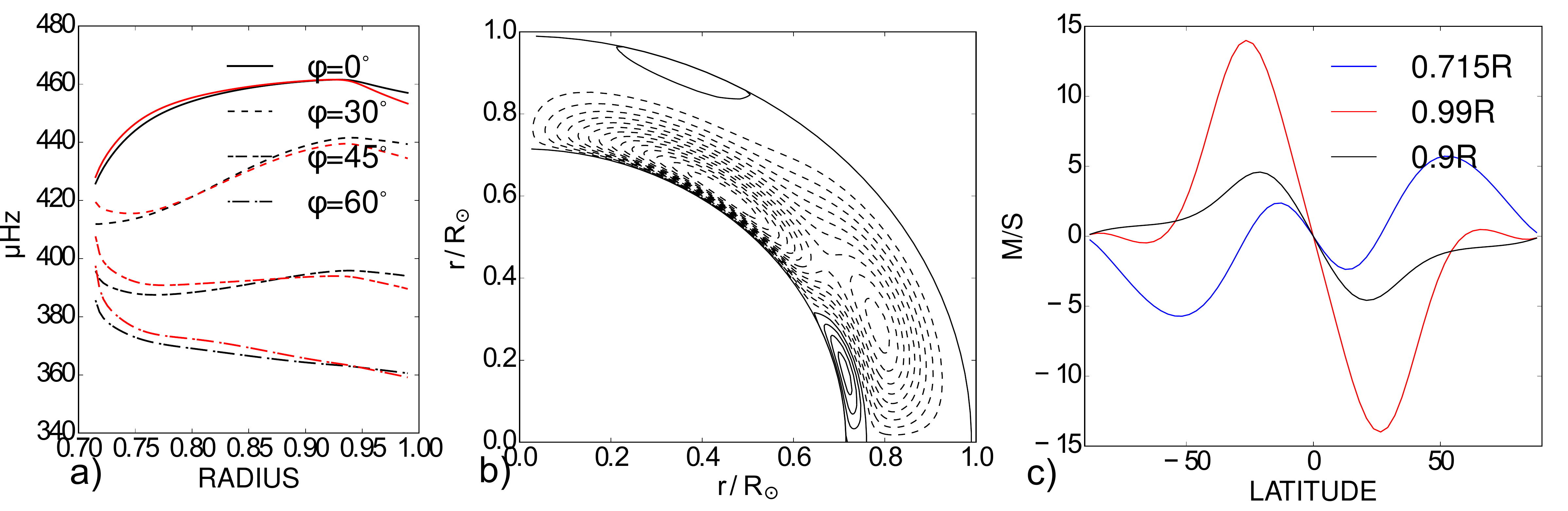}

\protect\caption{\label{figM2}Model M2: a) the radial profiles of the angular
velocity for different latitudes. The red color shows results for
model M1 for comparison; b) the stream function distribution, the dashed lines show
the counterclockwise circulation and the solid lines show the opposite
circulation;  c) velocity of the latitudinal component of the meridional flow
for different radii.}
\end{figure}

\begin{figure}
\includegraphics[width=0.95\linewidth]{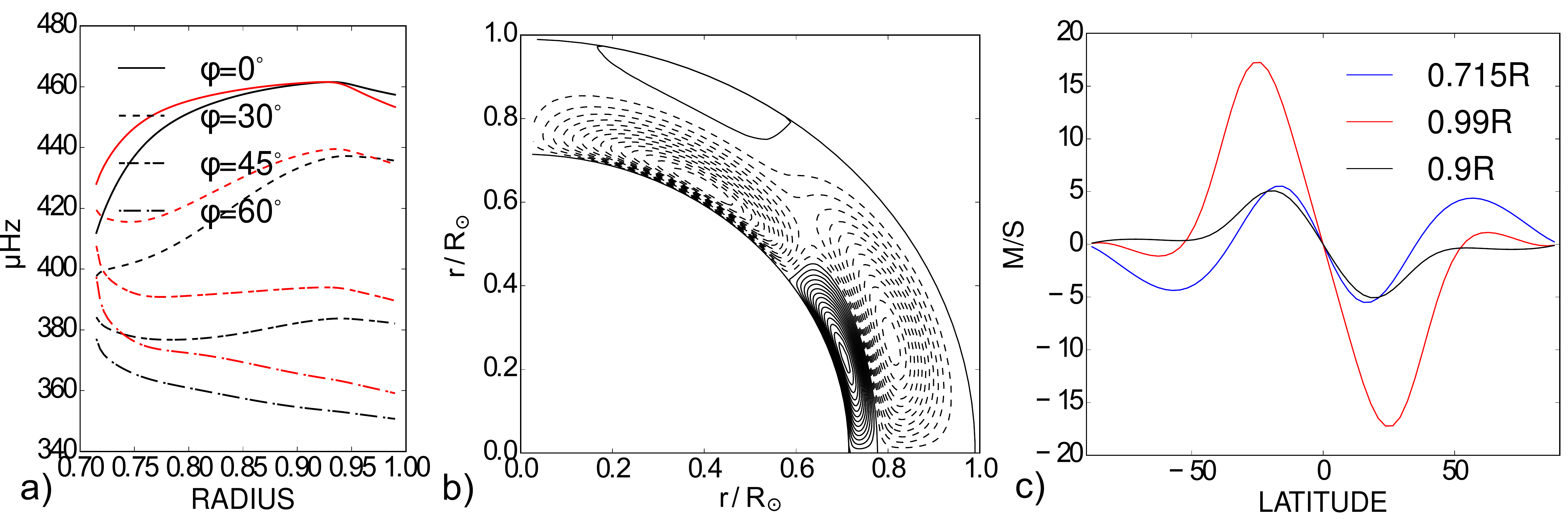}

\protect\caption{\label{figM3}The same as in Figure \ref{figM2} for the model M3.}
\end{figure}

Further increase of the inward angular momentum flux (via 
parameter $c_{\Lambda}$) results in amplification of the near equatorial
clockwise circulation cell. This is illustrated by  model M3 (Figure
\ref{figM3}).
 This model has a slightly stronger latitudinal
shear at the surface than  model M1. The amplitude of the meridional
flows in model M3 is larger than in  M2. The  flow speed  reaches
 about 18 m/s at the surface and about  5 m/s at the bottom of the convection zone. Model M3 
qualitatively preserves the angular velocity profile of model M1.

\section{Discussion and conclusions}

We investigated the mean-field models of the solar differential rotation
using a simple description of the turbulent $\Lambda$-effect. Our goal was to
study how the distributions of the non-dissipative angular momentum fluxes
affects the meridional circulation structure in the solar convection zone. The
study is motivated by the recent findings of helioseismology showing
the existence of a  double-cell meridional circulation pattern
\citep{Zhao13m}, and also  results of
the global 3D numerical simulations which often demonstrate that 
circulation structure can be multicellular (see, e.g., \citet{kap2011,miesch11,guer2013}. 

\begin{figure}
\includegraphics[width=0.9\linewidth]{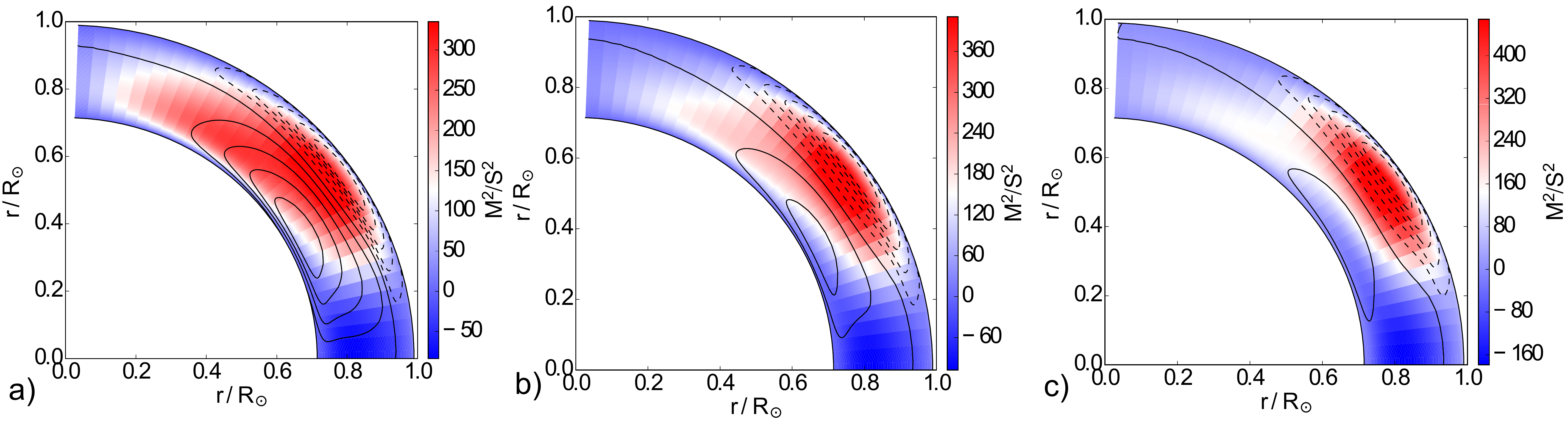}\protect\caption{The Reynolds stresses $T_{r\phi}$ (background color) and $T_{\theta\phi}$
(contours, in the same range as $T_{r\phi}$), for models: a) M1; b)
M2 ;c) M3. }

\end{figure}

Inspecting Eqs.(\ref{eq:trf},\ref{eq:V}) we see that the transport
of angular momentum due to the turbulent $\Lambda$- effect changes from outward
to inward near the equator. The near equatorial inward
flux is minimal in model M1 and it grows with the decrease of
 parameter $c_{\Lambda}$. The non-monotonic spatial dependence
of the angular momentum fluxes, and the extent of the near-equatorial region
occupied by the inward non-dissipative angular momentum flux  provides
conditions for destabilization of the Taylor-Praudman balance, which
results to the increasing complexity of the meridional circulation
pattern. 

Figure 5 illustrates the Reynolds stresses $T_{r\phi}$ and $T_{\theta\phi}$
for our set of models. In model M3, characterized by a triple-cell
circulation pattern, the outward turbulent angular momentum fluxes
are concentrated close to the surface in the mid-latitude of the solar convection
zone. Also, in model M3 the amplitude of the inward flux
near the  equator is about factor three larger  than in model M1. The
turbulent Reynolds stress tensor component $T_{r\phi}$
is symmetric about the equator and  $T_{\theta\phi}$ is anti-symmetric.
Figure 5c (model M3) shows a similarity with the global 3D
simulations results  presented by \citet{kap2011} (see Figures 4,5 for run
A6 in their paper). Thus the origin of the multicellular merdional circulation
pattern in their model can be explained by the latitudinal variations
of the $\Lambda$-effect. This question should be studied further. 

The rotation profiles in all three models are qualitatively similar, while
 model M1 is, probably, in the best agreement with the results of helioseismology
inversions of \citet{Howe2011JPh}m models M2 and M3 also have 
the radial profile of the angular velocity in qualitative agreement
with helioseismology. Thus, we conclude that the multicellular meridional
circulation structure can be explained by the mean-field models.
However our models do not include the solar tachocline. From the results of \citet{kit11},
we can guess that inclusion of the tachocline may affect the Taylor-Paudman
balance near the bottom of the convection zone and change the magnitude of the meridonal
circulation.

The turbulent part of the angular momentum transport is not well understood.
The mean-field theory of the $\Lambda$-effect \citep{KR93L} has
some issues that are rarely discussed in the literature. The theory
is constructed for forced isothermal turbulence rather than for turbulent
convection. Calculation of \citet{kle06} show that the
$\Lambda$-effect  functions $V^{(0)}$ and $V^{(1)}$ may have
dependence on the Coriolis
number that is different from the results of  \citet{KR93L}. Thus, theoretically, there
is some uncertainty in the description of the turbulent angular
momentum fluxes  in rotating stellar convection zones. These uncertainties should
be resolved using the global 3D numerical simulation and assimillation
of the observational data in theoretical models.

\textbf{Acknowledgments} VP thanks support of RFBR under grant 
15-02-01407 and the project II.16.3.1 of ISTP SB RAS.

%\bibliographystyle{/home/va/work/pap/aastex/3/apj}
%\bibliography{/home/va/work/pap/dyn}

\end{document}